\documentclass[twocolumn,superscriptaddress,showpacs,preprintnumbers,amsmath,amssymb,prl]{revtex4}

\usepackage[dvips]{graphicx}
\usepackage{dcolumn}
\usepackage{bm}
\newcommand{\msr}{$\mu$SR}

\newcommand{\lvo}{LiV$_2$O$_4$}

\newcommand{\ymn}{YMn$_2$}
\newcommand{\ysmn}{Y$_{1-x}$Sc$_x$Mn$_2$}

\begin{document}

\preprint{APS/123-QED}
\title{Quasi-One-Dimensional Spin Dynamics in $d$-Electron Heavy-Fermion Metal \ysmn}

\author{Masanori Miyazaki}
\affiliation{Department of Materials Structure Science, The Graduate University for Advanced Studies, Tsukuba, Ibaraki 305-0801, Japan}
\author{Ryosuke Kadono}
\affiliation{Department of Materials Structure Science, The Graduate University for Advanced Studies, Tsukuba, Ibaraki 305-0801, Japan}
\affiliation{Muon Science Laboratory and Condensed Matter Research Center, Institute of Materials Structure Science, High Energy Accelerator Research Organization (KEK), Tsukuba, Ibaraki 305-0801, Japan}
\author{Masatoshi Hiraishi} 
\affiliation{Department of Materials Structure Science, The Graduate University for Advanced Studies, Tsukuba, Ibaraki 305-0801, Japan}
\author{Tetsuya Masuda}
\affiliation{Muon Science Laboratory and Condensed Matter Research Center, Institute of Materials Structure Science, High Energy Accelerator Research Organization (KEK), Tsukuba, Ibaraki 305-0801, Japan}
\author{Akihiro Koda}
\affiliation{Department of Materials Structure Science, The Graduate University for Advanced Studies, Tsukuba, Ibaraki 305-0801, Japan}
\affiliation{Muon Science Laboratory and Condensed Matter Research Center, Institute of Materials Structure Science, High Energy Accelerator Research Organization (KEK), Tsukuba, Ibaraki 305-0801, Japan}
\author{Kenji~M. Kojima}
\affiliation{Department of Materials Structure Science, The Graduate University for Advanced Studies, Tsukuba, Ibaraki 305-0801, Japan}
\affiliation{Muon Science Laboratory and Condensed Matter Research Center, Institute of Materials Structure Science, High Energy Accelerator Research Organization (KEK), Tsukuba, Ibaraki 305-0801, Japan}
\author{\\ Teruo Yamazaki}
\affiliation{Department of Material Science and Engineering, Kyoto University, Kyoto, 606-8501, Japan}
\author{Yoshikazu Tabata}
\affiliation{Department of Material Science and Engineering, Kyoto University, Kyoto, 606-8501, Japan}
\author{Hiroyuki Nakamura}
\affiliation{Department of Material Science and Engineering, Kyoto University, Kyoto, 606-8501, Japan}

%

\begin{abstract}

Slow spin fluctuations ($\nu<10^{12}$ s$^{-1}$) observed by the muon spin relaxation technique in \ysmn\ exhibits a power law dependence on temperature ($\nu\propto T^\alpha$), where the power converges asymptotically to unity ($\alpha\rightarrow1$) as the system moves away from spin-glass instability with increasing Sc content $x$.  
This linear $T$ dependence, which is common to that observed in \lvo, is in line with the prediction of the ``{\sl intersecting} Hubbard chains" model for a metallic pyrochlore lattice, suggesting that the geometrical constraints to $t_{2g}$ bands specific to the pyrochlore structure serve as a basis of the $d$-electron heavy-fermion state.

\end{abstract}

\pacs{75.25.+z, 75.50.Lk, 76.75.+i}
\keywords{geometrical frustration, heavy fermion, intersecting Hubbard chains, quasi-1D spin dynamics, muon spin rotation}
\maketitle

Geometrical frustration in electronic degrees of freedom such as spin, charge, and orbit, which is often realized on stages of highly symmetric crystals, has been one of the major topics in the field of condensed matter physics. In particular, numerous studies of local electronic systems (i.e., insulators) have revealed a wide variety of interesting phenomena associated with geometrical frustration, e.g., incommensurate order with long periodicity, successive unfolding of self-organized structures at different energy scales, nontrivial effects of strong fluctuation originating from the degeneracy of states over a macroscopic scale, weakened critical divergence, and associated new universality class of criticality.

Although a similar situation may be speculated on, little is actually known about the effect of geometrical correlation on {\sl itinerant} electron systems.  As one of few such examples, the heavy-fermion behavior in a cubic vanadium spinel, \lvo, has received considerable interest \cite{Kondo:97,Urano:00}, where such a local electronic correlation specific to the highly symmetric pyrochrore structure may be of direct relevance to the formation of heavy-quasiparticle (QP) state. 

The heavy QP mass ($m^*$) is phenomenologically understood to come from the sharp increase in the electronic density of states near the Fermi surface and associated flattening of band dispersion [$dN(E_{\rm F})/dE\propto m^*\rightarrow\infty$].  In rare-earth compounds, such enhancement is induced by the conversion of local $f$-electron degrees of freedom into $N(E)$ by the Kondo effect, which is observed as a peak structure of $N(E)$ near $E_{\rm F}$.
While such a structure in $N(E)$ has been suggested in \lvo\ by photoemission spectroscopy \cite{Shimo:06}, the dynamical fluctuation of seemingly ``local" vanadium magnetic moments associated with the heavy-QP state has been inferred from muon spin rotation and relaxation (\msr) measurements \cite{Koda:04,Koda:05}.  The latter indicates that the conventional Kondo coupling (that virtually eliminates ``local" spins over a time scale longer than $\nu_{\rm ex}^{-1}=h/J_{cf}\sim10^{-13}$--$10^{-14}$ s, where $J_{cf}$ is the exchange energy between conduction electrons and $f$ electrons) is not in effect, thereby suggesting different origin of the heavy QP state.  

This situation naturally turns our interest to yittrium manganite (\ymn), an intermetallic Laves phase (C15-type) compound that is the first example of transition-metal systems in which a heavy-QP state is observed.  It comprises a three-dimensional network of corner-shared tetrahedra with Mn ions at their corners, providing a stage equivalent to the cubic pyrochlore lattice. Although \ymn\ exhibits magnetic order with complex helical modulation and a large volume expansion below $T_N\sim$100 K \cite{Ballou:87}, it remains in a paramagnetic state under hydrostatic pressure ($\ge0.4$ GPa) \cite{Oomi:87} or upon the substitution of Y by Sc (\ysmn, with $x\ge0.03$) \cite{Nakamura:88} which is also accompanied by a large increase in QP mass ($m^*\sim$15 times the band mass) as inferred from the electronic specific heat \cite{Wada:89,Fisher:93}.  Despite studies focused onto identifying the origin of the heavy-QP state, the issue still stands as a major challenge that has increased in importance with the succeeding discovery of \lvo.

Here, we report on the spin dynamics of \ysmn\ studied by the muon spin relaxation (\msr) technique at various Sc contents $x$. It is shown that Mn spin fluctuation persists over a low frequency range ($\nu<10^{12}$ s$^{-1}$), which is characterized by a power law temperature dependence, $\nu\simeq c\cdot T^\alpha$. The power $\alpha$ asymptotically approaches unity with increasing $x$ ($\propto$ chemical pressure), while $\alpha\sim2$ as $\nu$ shows tendency of rapid slowing down toward a quasistatic spin-glass state near $x=0.03$.  The presence of such linear $T$ dependence of spin fluctuation  strikingly resembles that in the case of \lvo, which is understood as a characteristic property of spin-spin correlations for the intersecting Hubbard chains as a model of the pyrochlore lattice \cite{Lee:03}. This implies the crucial role of $t_{2g}$ orbitals as one-dimensional chains that are under a strong geometrical constraint of pyrochlore lattice structures, and further suggests the dimensional crossover due to coupling between these chains as one of the origins of the heavy-fermion state \cite{Fujimoto:02}.

Polycrystalline samples of \ysmn\ with various Sc contents ($x=0.03$, 0.05, 0.07, and 0.08, as prepared) were grown from melts in an argon arc furnace followed by annealing, where the details of sample preparation are described elsewhere \cite{Nakamura:88}.  
The magnetization and specific heat measurements were performed on the samples grown under the same condition as those for \msr, where the thermal relaxation method was used for the latter measurements. 

\begin{figure}[t]
	\centering
	 \includegraphics[width=0.9\linewidth,clip]{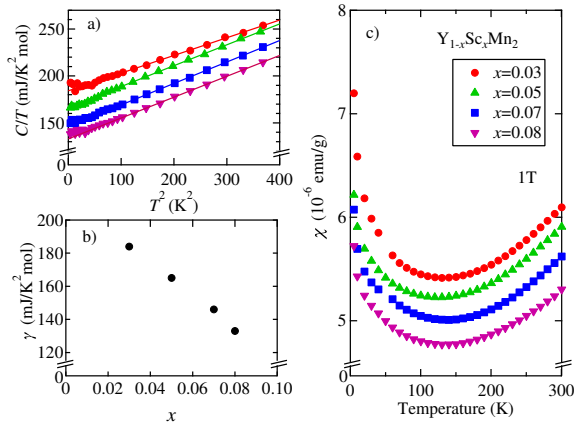}
	\caption{(Color online) (a) Specific heat divided by temperature ($C/T$) vs $T^2$ in \ysmn\ with $x=0.03$, 0.05, 0.07, and 0.08. (b) Electronic specific heat coefficient ($\gamma$) vs $x$ deduced by curve fitting of data shown in (a). (c) Magnetic susceptibility (measured at 1 T) vs temperature. } \label{gamma}
\end{figure}

As shown in Fig.~\ref{gamma}a, the specific heat divided by temperature ($C/T$, plotted against $T^2$) exhibits a systematic variation with Sc content, from which, as shown in Fig.~\ref{gamma}b, one can deduce the electronic specific heat coefficient $\gamma$ by curve fitting using the relation $C/T=\gamma +a\cdot T^2$, where the second term represents the contribution of phonons. The obtained values are consistent with earlier literature values (reported $\gamma$ values for $x=0.03$ and 0.08), and much greater than that predicted by band calculation ($\sim9$ mJ/K$^2$mol), where $m^*$ is close to that in the case of UPt$_3$ \cite{Wada:89,Yamada:87,Asano:88}.  A similar mass enhancement was observed in \ymn\ under hydrostatic pressure to suppress the structural transition below $\sim$100 K, indicating that the Sc substitution corresponds to chemical pressure \cite{Fisher:93,Ballou:94}. The general trend of reduced 
$\gamma$ as well as the magnetic susceptibility $\chi$ (Fig.~\ref{gamma}c) at a greater $x$ is qualitatively in line with the reduction in $N(E_{\rm F})$ due to enhanced band width. 
The development of a Curie-Weiss-like behavior in $\chi$ at lower temperatures (although it is weak, as implied by a large offset of vertical axis in Fig.~\ref{gamma}c) is similar to that in the case of \lvo, whose possible origin is discussed below. 

Conventional \msr\ measurements were performed on the M20 beamline of TRIUMF, Canada. A \msr\ apparatus furnished with a superconducting magnet was used to measure the time-dependent positron decay asymmetry at a longitudinal field (LF) of up to 5~T.   The samples were loaded on a He gas-flow cryostat with special precaution to avoid exposure to air that would deteriorate sample quality.  A magnetic field was applied parallel to the initial muon spin direction $\vec{P}_\mu(0)$, which was also parallel to the muon beam direction ($\hat{z}$-axis). The time-dependent muon polarization [$G_z(t)=\hat{z}\cdot \vec{P}_\mu(t)$] was monitored by measuring the decay-positron asymmetry along the $\hat{z}$-axis  [$A(t)=A_0G_z(t)$, with $A_0\simeq0.2$ being the initial asymmetry] . 

\begin{figure}[tb]
\begin{center}
\includegraphics[width=0.8\linewidth]{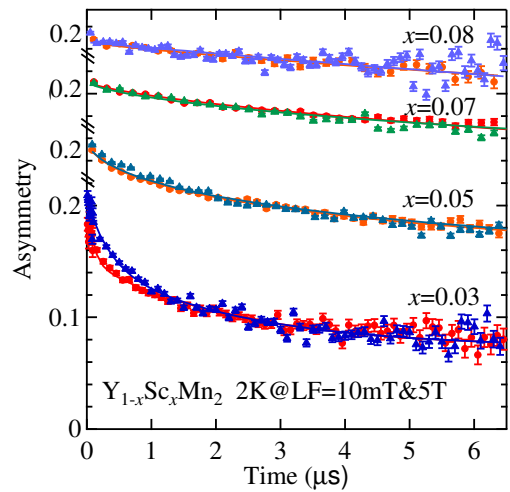}
\caption{(Color online)
Typical examples of \msr\ spectra [time-dependent asymmetry, $A(t)$] observed in \ysmn\ at 2 K under two different longitudinal fields [$H_0=10$ mT (triangles) and 5 T (circles)], where spectra for $x\ge0.05$ are shifted vertically by 0.05 with each $x$ for clarity.  Solid curves show best fits using the function described in the text. 
}
\label{tspec}
\end{center}
\end{figure}

\begin{figure}[tb]
\begin{center}
\includegraphics[width=0.34\textwidth]{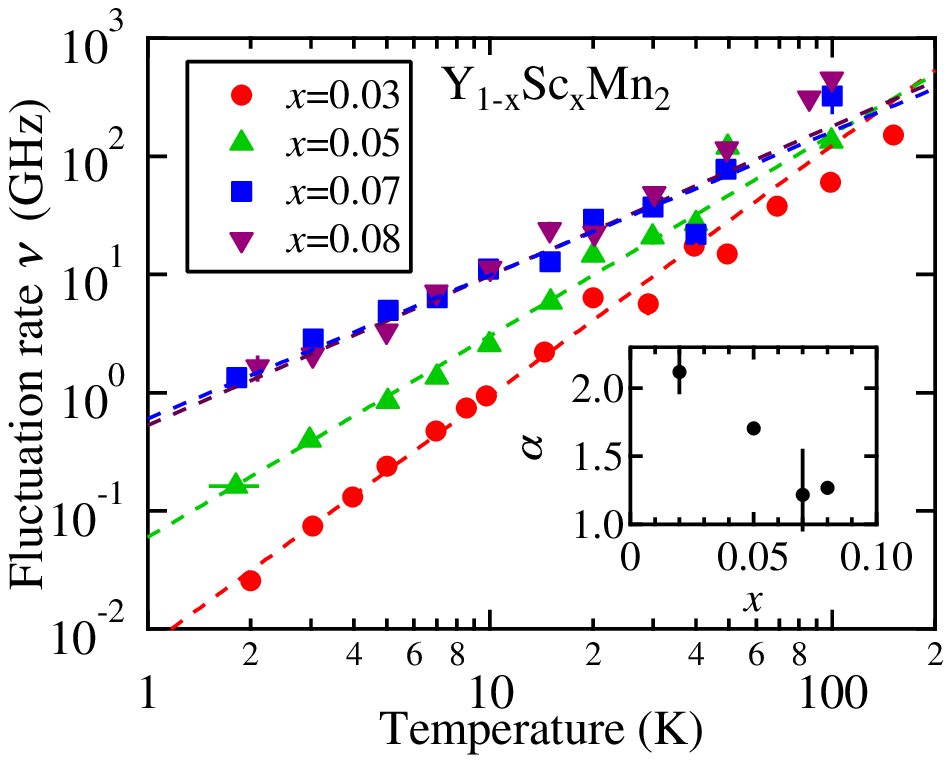}
\includegraphics[width=0.13\textwidth]{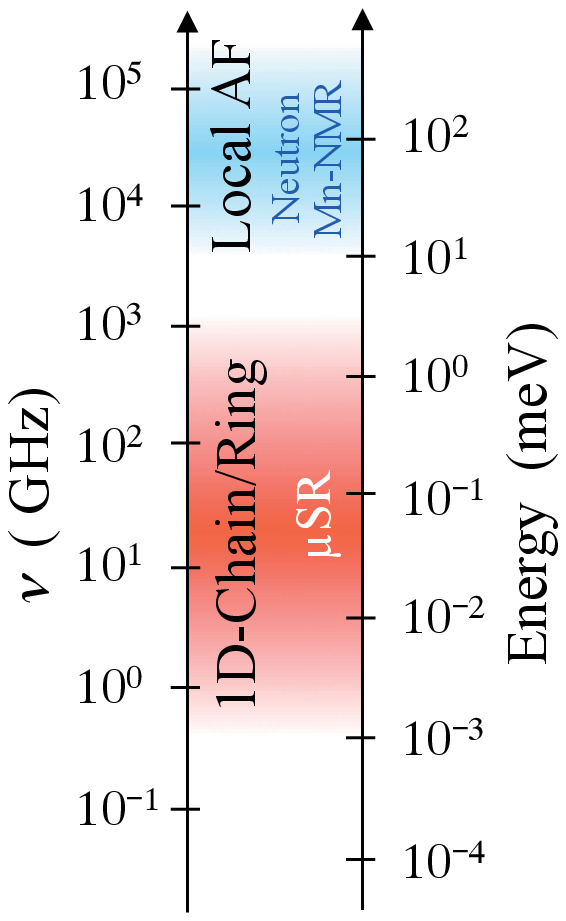}
\caption{(Color online)
Spin fluctuation rate ($\nu$) as a function of temperature and Sc content ($x$)  in \ysmn.  Solid lines are results of curve fitting using a power law ($\nu\propto T^\alpha$). Inset: $\alpha$ obtained by curve fits vs $x$.  Right: characteristic range of $\nu$ and corresponding energy scale  (see text).
}
\label{nutemp}
\end{center}
\end{figure}

Figure \ref{tspec} shows some examples of LF-\msr\ spectra at low temperatures ($\simeq2$ K), where the depolarization rate ($\lambda$) decreases with increasing Sc content $x$ and tends to approach an asymptotic value (as it exhibits little change between $x=0.07$ and 0.08).  
It is also noticeable that $\lambda$ is mostly independent of the magnitude of external 
field ($H_0$) for $x\ge0.05$, while it shows a slight variation with $H_0$ for $x=0.03$.  These features can be readily understood within the conventional model of metals, which shows $\lambda$ under fluctuating hyperfine fields as
\begin{equation}
\lambda(\nu,H_0)\simeq \frac{2\delta_\mu^2\nu}{\nu^2+\gamma_\mu^2 H_0^2}\cdot k_BT\chi,\label{rfd}
\end{equation}
where $\delta_\mu$ is the hyperfine field exerting on muons, $\nu$ is the fluctuation rate of $\delta_\mu$, and $\gamma_\mu$ is the muon gyromagnetic ratio ($=2\pi\times135.53$ MHz/T). Equation (\ref{rfd}) is valid when $\nu\gg\delta_\mu$. Assuming that muons are located at the 16$c$ site (or its vicinity, as reported in earlier literature including those on hydrogen in \ymn\ \cite{Mekata:00,Hartmann:90,Fujiwara:87}), $\delta_\mu$ is calculated from the muon-Mn dipolar tensor to yield $\delta_\mu/\mu_B=$ 36 MHz/$\mu_B$, which we adopt as a constant for the rest of the analysis.   This is partly justified from the fact that, in the sample with $x=0.03$, the value is in good agreement with that estimated from the field dependence of $\lambda$  [$\delta_\mu=41(2)$ MHz] and that the Mn moment size suggested from neutron scattering is $\sim$1$\mu_B$\cite{Shiga:88}, while the independence of $\delta_\mu$ against $x$ remains as a reasonable hypothesis.  We also note that eq.~(\ref{rfd}) has been successfully applied to various types of magnetisms including that of quasi-one-dimentional compounds \cite{Yamauchi:10}. Equation (\ref{rfd}) is modified to yield the fluctuation rate $\nu\simeq \delta_\mu^2k_BT\chi/\lambda+\sqrt{(\delta_\mu^2k_BT\chi)^2/\lambda^2+\gamma_\mu^2 H_0^2}$, from the experimental values of $\lambda$ and $\chi$. As shown below (see Fig.~\ref{nutemp}), the deduced $\nu$ turns out to be always greater than $\delta_\mu$, indicating the valid use of eq.~(\ref{rfd}).  

Least-squares curve fits were attained to deduce $\lambda$ from \msr\ time spectra using 
\begin{equation}
AG_z(t)=A_0[(1-a_p)\exp\{-(\lambda t)^\beta\}+a_p],\label{gzt}
\end{equation}
with the power $\beta$ and a constant term $a_p$ as additional parameters. Here, we introduce the stretched exponential decay to reproduce the possible effect of randomness due to Sc substitution and the associated deviation of spin dynamics from that described by the simple spin correlation model with a single value of $\delta_\mu$ and/or $\nu$ at a given temperature \cite{Mekata:00,Johnston:05}. The presence of the term $a_p$ is clearly inferred in the case of $x=0.03$ from the leveling off of the time spectra for $t\ge4$ $\mu$s (see Fig.~\ref{tspec}, needed for $x=0.05$ as well). Although its origin is unclear at this stage, the nonzero $a_p$ suggests more weight for $\nu\gg\delta_\mu$ in the spectral density $P(\nu)$.  Apart from such ambiguity, we note that excellent fits were obtained in all of the cases with $a_p$ fixed to the values deduced at 2 K  [{\it i.e.}, independent of temperature, with $a_p=0.348(2)$ and 0.277(2) for $x=0.03$ and 0.05, respectively, and $a_p=0$ for other cases].  It also turned out that the fits were insensitive to $\beta$ for $x=0.08$ at higher temperatures; therefore $\beta$ was fixed to the value deduced at 2 K [$\beta=0.678(4)$]. Solid curves in Fig.~\ref{tspec} represent the best fits obtained under these conditions.

 \begin{figure}[tb]
\begin{center}
\includegraphics[width=0.9\linewidth]{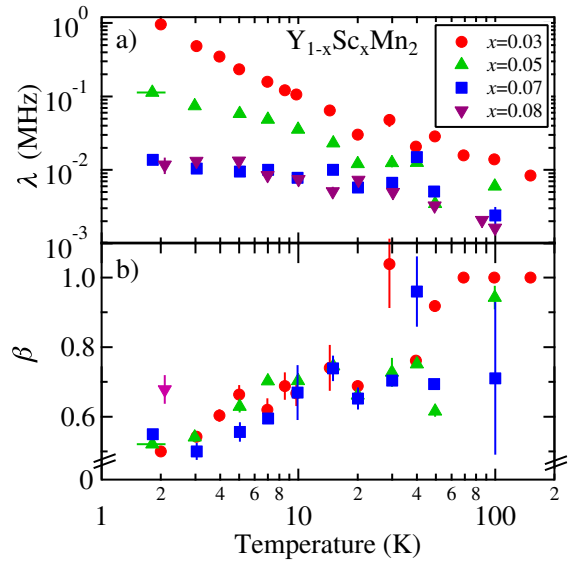}
\caption{(Color online)
Temperature dependences of (a) depolarization rate ($\lambda$) and power ($\beta$) in \ysmn\ obtained by curve fitting of $A(t)$ for $x=0.03$ (circles), 0.05 (triangles), 0.07 (squares), and 0.08 (inverted triangles), respectively.  $\beta$ for $x=0.08$ is fixed at 2 K (only which is shown here). 
}
\label{tparams}
\end{center}
\end{figure}

The temperature dependences of the parameters deduced from curve fits are summarized in Fig.~\ref{tparams}.  Although $\beta$ varies only slightly with $x$, $\lambda$ exhibits a clear tendency of becoming less dependent on temperature ($T$) with increasing $x$, showing a $T$ dependence similar to that of $\chi(T)$ for $x\ge0.07$. Note that this behavior is apparently different from that  expected from the Korringa relation ($\lambda\cdot T^{-1}=const.$) for normal metals. 
Considering the way how $\lambda$ depends on $\nu$, and $\chi$ in eq.~(\ref{rfd}), this would mean that $\nu$ becomes linearly dependent on $T$
with increasing $x$.  
Meanwhile, the behavior of $\lambda$ for $x\rightarrow0.03$ is understood as the freezing of the Mn spin fluctuation, as the transition to the quasistatic spin-glass state occurs in the sample with $x=0.03$ below $T_g\simeq 3$ K (where $T_g$ is defined as the peak muon depolarization rate under LF=10 mT)\cite{Mekata:00}.  
The behavior of $\lambda$ observed for $x\ge0.07$ shows a distinct similarity to that in \lvo\ \cite{Koda:04}.

Figure \ref{nutemp} shows the $T$ dependence of $\nu$ deduced from $\lambda$ using eq.~(\ref{rfd}), where $\nu$ is in the range of $10^0$--$10^2$ GHz ($10^9$ --$10^{11}$ s$^{-1}$) in the sample with $x\ge0.07$ over the observed temperature range of $10^0$--10$^2$ K, while it shows a steeper reduction with decreasing temperature for $x\le0.05$. Although the use of a stretched exponential decay in eq.~(\ref{gzt}) prevents $\nu$ from being simply interpreted as a mean when $\beta<1$, $\nu$ serves as a ``characteristic frequency" that describes the spin dynamics on the basis of eq.~(\ref{rfd}) \cite{Johnston:05}.
Solid lines are obtained by curve fitting using a power law, $\nu=c\cdot T^\alpha$, with $c$ and $\alpha$ being free parameters. [These are only for data below 20-40 K, which  are relatively free of systematic uncertainty coming from correlation among fitting parameters ($\nu$ and $\beta$) in the preceding curve fits of the time spectra; see the scattering of parameters at higher temperatures in Fig.~\ref{tparams}] As shown in the inset of Fig.~\ref{nutemp}, $\alpha$ exceeds 2 in the case of $x=0.03$, whereas it approaches unity for $x\ge0.07$. 

It would be worth stressing that nuclear magnetic resonance (NMR) and inelastic neutron scattering (INS) studies of the paramagnetic phase of \ysmn\ conducted thus far are mostly concerned with the relatively high energy part of spin dynamics,  where they have demonstrated the presence of antiferromagnetic (AF) correlation with a characteristic frequency scale of $\nu_{\rm AF}\simeq10^{13}$--$10^{14}$ s$^{-1}$ (see Fig.~\ref{nutemp}, right) \cite{Nakamura:01,Zheng:99,Deportes:87,Shiga:88,Ballou:96}. In particular, a strong hyperfine field exerting on $^{55}$Mn nuclei  [corresponding to $\delta_\mu\simeq1.2$ GHz in eq.~(\ref{rfd})] drives the sensitive range of NMR up to such high frequencies \cite{Zheng:99}.
Interestingly, the latest INS study of a single-crystalline sample ($x=0.03$) revealed that the intensity centered at approximately ${\bf Q}_0=(1.25,1.25,0)$ (in reciprocal lattice units) exhibits anisotropic broadening along the Brillouin zone boundary, which is interpreted to be due to the existence of the degeneracy of states associated with geometrical frustration \cite{Ballou:96}.  Although this might be  reminiscent of low-energy spin dynamics, the details are yet to be clarified.

According to a theoretical investigation of intersecting Hubbard chains as a model of  pyrochlore sublattice in \lvo, the low-energy part of spin dynamics is predicted to exhibit a linearly $T$-dependent relaxation rate in the spin-spin correlation \cite{Lee:03}. The observed behavior of $\nu$ in \ysmn\ for a greater Sc content $x$ (where the system is far from spin-glass instability) is perfectly in line with the above prediction, suggesting that the $t_{2g}$ orbitals associated with Mn atoms retain their one-dimensional (1D) character over the relevant temperature (energy) range.



Here, we ought to point out that the spin fluctuation rate in \lvo\ obtained by earlier \msr\ measurements ($\nu_D$ in ref.~[\onlinecite{Koda:04}]) was derived from the corresponding muon depolarization rate ($\lambda_D$) using a theoretical model that would have been valid only for the limit of localized vanadium  moments: re-evaluation using eq.~(\ref{rfd}) for the metallic state indicates that $\nu_D$ exhibits a linear $T$ dependence \cite{Kadono:11}, which is perfectly in line with the behavior of the relaxation rate ($\Gamma_Q\propto T$ for $T<10^2$ K) observed over the low-energy region of the INS spectrum \cite{Lee:01}.  This implies that both \msr\ (${\bf q}$-averaged) and INS have been probing similar parts of the spin fluctuation spectrum of \lvo, which is common to the present case of \ysmn. 

Among the many theoretical models for the microscopic origin of heavy fermion state in \lvo, that by Fujimoto regards the quasi-1D character of the $t_{2g}$ bands associated with the pyrochlore lattice (consisting of intersecting chains of $t_{2g}$ orbitals) as an essential basis for the description of electronic state, as it is expected that the hybridization between the 1D bands will be strongly suppressed owing to the geometrical configuration \cite{Fujimoto:02}. The model incorporates the hybridization as a perturbation to the 1D Hubbard bands, which yields an energy scale ($T^*$) that characterizes the dimensional crossover from 1D to 3D as the Fermi-liquid state develops with decreasing temperature below $T^*$.  The calculated specific heat coefficient taking account of the latter as the leading correction to the self-energy yields a large value, consistent with the experimentally observed ones.  The progression of hybridization also induces the enhancement of the 3D-like spin correlation that would appear as the enhancement of uniform susceptibility, while the spin fluctuation is dominated by the staggered component of 1D Hubbard chains.    The increase in $\chi$ in \ysmn\ at low temperatures (which is common to the case of \lvo) may be understood as the manifestation of such a dimensional crossover with $T^*\simeq10^2$ K, while the increase in $\chi$ with $T$ for $T>10^2$ K is typically associated with the low-dimensional character of AF spin correlation.  The quasi-1D character of the low-energy spin fluctuation preserved  below $T^*$ coexisting with the enhanced $\chi$ comprises strong evidence of such a scenario.

Finally, we note that the weak temperature dependence of $\beta$ observed in Fig.~\ref{tparams}b might be attributed to the dissipative character of excitation energies involved in spin dynamics induced by the random potential of Sc substitution \cite{Philips:96}. To eliminate such an extrinsic effect, a  \msr\ experiment on \ymn\ under hydrostatic pressure is currently under consideration.

In summary, we have shown by \msr\ study that the low-energy spin dynamics in \ysmn\ is characterized by a linear $T$ dependence of fluctuation rate that is explained by the theoretical model of intersecting 1D Hubbard chains.  This property is common to another $d$-electron heavy-fermion system, \lvo, and indicates that a geometrically constrained $t_{2g}$ band is the primary stage for the formation of heavy quasiparticles with  
the 1D-to-3D dimensional crossover as a possible mechanism.

 We would like to thank the TRIUMF staff for their technical support during the $\mu$SR experiment. This work was partially supported by the KEK-MSL Inter-University Program for Oversea Muon Facilities and by a Grant-in-Aid for Creative Scientific Research on Priority Areas from the Ministry of Education, Culture, Sports, Science and Technology, Japan.

\end{document}